\documentclass[aps,prl,reprint,superscriptaddress,nofootinbib,longbibliography]{revtex4-1}

\usepackage{amssymb,amsmath,bm,natbib}
\usepackage[dvipsnames]{xcolor}
\usepackage[colorlinks = true,
            linkcolor = Maroon,
            urlcolor  = Maroon,
            citecolor = Maroon]{hyperref}
\usepackage{microtype}
\usepackage{graphicx}
\usepackage{tikz}
\usetikzlibrary{decorations.pathmorphing}
\usepackage{xspace}
\usepackage{bm}

\newcommand{\be}{\begin{equation}}
\newcommand{\ee}{\end{equation}}

\newcommand{\spacy}{\mkern1mu} 

\begin{document}

\title{Quantum limit to transparency} 

\author{Angelo Esposito}
\affiliation{Dipartimento di Fisica, Sapienza Universit\`a di Roma, Piazzale Aldo Moro 2, I-00185 Rome, Italy}
\affiliation{INFN, Sezione di Roma, Piazzale A. Moro 2, I-00185 Rome, Italy}

\author{Alberto Nicolis}
\affiliation{Department of Physics, Center for Theoretical Physics, Columbia University, 538W 120th Street, New York, NY, 10027, USA}

\author{Federico Piazza}
\affiliation{Aix Marseille Univ, Universit\'e de Toulon, CNRS, CPT, Marseille, France}

\begin{abstract}
\noindent We argue that no material can be made completely transparent to light as long as it has a nontrivial index of refraction. The quantum decay of photons into phonons is an ineliminable source of attenuation. This survives even once all other excitations have been gapped away, impurities have been removed, and the temperature has been set to zero. Restricting for simplicity to the case of a superfluid, we determine  via effective field theory methods the corresponding low-frequency contribution to the imaginary part of the index of refraction. This scales as $\omega^4$ and induces a nonanalyticity of the real part for $\omega \rightarrow 0$. We also extend our result to large coherent electromagnetic fields, such as those of a laser, showing that the effect can receive strong nonlinear enhancements. Our result is a genuine quantum lower bound to the transparency of a medium.
\end{abstract}


\maketitle


\noindent\emph{\textbf{Introduction}} --- How transparent can a material be?
Clearly, if a material does not interact with light at all---say,  a hypothetical medium made of dark matter---there is no limit to its transparency. So, let's qualify the question and make it more meaningful: Assuming we have a material with a nontrivial refraction index, $n \neq 1$, is there a fundamental limit to how transparent it can be?  

Departures from transparency are parametrized by the imaginary part of $n$, which is a measure of how easily  the energy of light can get dissipated in the material. So, to make a material more transparent, one needs to minimize the sources of dissipation: {\em i)} The electromagnetic field couples to the electric charge, and so we can start by demanding that in the material there are no {\em gapless} charged excitations, which is the definition of an insulator: a photon with frequency  below the insulator's gap does not have enough energy to produce charged excitations. This closes off a possible channel of dissipation al low enough frequencies.
{\em ii)} However, at nonzero temperature all excitations---even gapped ones---are thermally excited. A photon traveling in a thermal bath of  excitations will be scattered by them, and some of its energy will get dissipated into this thermal bath, yielding a non-zero imaginary part of the refraction index.\footnote{The attenuation length due to the interaction between photons and thermal density waves was first computed by Einstein~\cite{einstein1910theorie,landau2013electrodynamics}.} So, a second requirement to minimize dissipation is that the the temperature be as low as possible, ideally  zero. {\em iii)} Finally, impurities (or more general sources of disorder) generically lead to dissipation, because our photon can scatter off them. So, ideally we want to be dealing with an absolutely pure material, such as a perfect crystal. 

In conclusion, it appears that if we have a pure insulator at zero temperature, at frequencies below the gap the index of refraction has exactly vanishing imaginary part. We claim that this is {\em false}. The point is that all media have internal gapless excitations, corresponding to acoustic phonons, which represent an irreducible source of dissipation. In particular, a photon can decay into a slightly lower energy one and a phonon. 
The presence of these excitations in the spectrum of material media is an unavoidable consequence of Goldstone's theorem, as they are the Goldstone bosons associated with the spontaneous breaking of spacetime symmetries, which is unavoidable in any material~\cite[e.g.,][]{Leutwyler:1996er,Nicolis:2015sra,Esposito:2026diu}.
As an example, in this letter we study this phenomenon in a zero-temperature superfluid. 

This is a quantum mechanical process, which can be understood in different ways. The simplest one is arguably the fact that the incoming photon can scatter off the fundamental constituents of the material, consequently exciting their collective motion, which is described by a quantum oscillator. From a more field-theoretic viewpoint, this can be regarded as the scattering of the incoming photon off the virtual phonons that populate the ground state of the material\footnote{From this viewpoint, our effect is related to stimulated Brillouin scattering, which is usually phrased in terms of electrostriction and radiation pressure \cite{Brillouin}.}. And so, in a sense, it is the quantum, zero-temperature version of the thermal-bath source of dissipation alluded to above. In fact, at low enough temperatures the thermal bath of excitations will be dominated by the acoustic phonons, since these are the only gapless degrees of freedom in an insulator. From this viewpoint then, our effect is due to the surviving of acoustic phonon fluctuations at zero temperature, with the thermal ones continuously morphing into the quantum ones as the temperature vanishes. This intuition will be confirmed in the coming sections.

To get the imaginary part of $n(\omega)$, we will use quantum effective field theory. 
We will show that our effect is the leading source of dissipation at low frequencies in a zero-temperature insulator, proving that
\begin{align} \label{eq: Im n}
	{\rm Im} \, n(\omega) \to C \, \omega^4 \quad {\rm as} \quad \omega \to 0 \, ,
\end{align}
where the constant $C$ is calculable in terms of the zero-frequency limit of $n(\omega)$ and its dependence on the material's density. 
Moreover, in quantum field theory imaginary parts and nonanalyticities usually go together. Indeed, in our case, associated with the imaginary part of $n$, there is a log-running of the $\omega^4$ term in the low-frequency expansion of the {\em real} part of $n(\omega)$: 
\begin{align} \label{eq: Re n}
	{\rm Re} \, n(\omega) = n_0 + n_2 \, \omega^2 - \frac{C}{\pi} \, \omega^4 \log \left(\frac{\omega^2}{\omega^2_0}\right) + \dots \, ,
\end{align}
where the constants $n_0$, $n_2$, and $\omega_0$ are not calculable within the effective field theory, but the coefficient $C$ is exactly the same as above. Eqs.~\eqref{eq: Im n} and \eqref{eq: Re n}, with the value for $C$ that we will compute below, are robust predictions of  effective field theory.

Finally, we will also show that, in the case of large coherent electromagnetic fields, additional nonlinear effects contribute to the fields' dissipation, and that, in some phenomenologically relevant regimes, these dominate over their incoherent counterpart.

\vspace{0.5em}

\noindent\emph{Conventions:} We employ a ``mostly minus'' signature for the spacetime metric, and natural units, $c=\hbar=\epsilon_0=1$.

\vspace{1em}


\noindent\emph{\textbf{The effective field theory}} ---
Consider a neutral superfluid. Its low-energy dynamics can be described by a Lorentz-scalar field $\psi(x)$ that transforms under a shift symmetry~\cite[e.g.,][]{Son:2002zn,Nicolis:2015sra}, $\psi(x) \to \psi(x) + {\rm constant}$, so it can appear in the effective action only through derivatives. In the absence of electromagnetic fields the action reads
\begin{align} \label{eq: phonon Lagrangian}
	{\cal L}= P(X ) + \dots \; , \qquad \text{ with } \; X=(\partial \psi)^2 \, ,
\end{align}
where $P$ is the pressure, and the dots stand for higher derivative terms, which we can neglect at low frequencies and wavevectors. On the ground state at constant chemical potential $\bar \mu$, the field $\psi$ takes the value $\langle\psi(x)\rangle = \bar \mu \, t$; we thus see that the function $P$ above is simply  the pressure as a function of the (squared) chemical potential, and is thus completely determined by the equation of state. 

Excitations above the ground state are described by fluctuations in the $\psi$ field: $\psi(x) = \bar \mu \big(t + \pi(x) \big)$.
The quanta of $\pi(x)$ are the superfluid phonons. Their free Lagrangian, which we get by expanding Eq.~\eqref{eq: phonon Lagrangian} up to quadratic order in $\pi$, reads
\begin{align} \label{L pi}
	{\cal L}_\pi \supset \frac{1}{2} \frac{w}{c_s^2} \left[ \dot \pi^2 - c_s^2 \big(\bm \nabla \pi\big)^2 \right] \,,
\end{align}
where $w = \rho + p$ and $c_s$ are the ground state's enthalpy density and sound speed,
\begin{align}
	w = 2 \spacy \bar\mu^2 \bar P' \, , \qquad  c_s^2 = \frac{\bar P'}{\bar P' + 2 \spacy \bar \mu^2 \bar P'' } \,.
\end{align}
(All barred quantities, from now on, are evaluated at the background value $\bar \mu$.)

The shift symmetry of the scalar field constrains also the coupling of the superfluid to an external electromagnetic field $A_\mu$~\cite{Caputo:2019xum}. The fact that our superfluid is neutral means that it cannot couple directly to $A_\mu$, but only to its gauge invariant field strength $F_{\mu\nu}$, and derivatives thereof. Moreover, although Lorentz invariance is broken inside the material, its breaking is {\em spontaneous}---it comes from $\psi$ having a Lorentz-breaking background configuration, $\langle\psi\rangle \propto t$. So, using Lorentz and shift symmetry on $\psi$, we see that to lowest order in derivatives and up to quadratic order in the electromagnetic field, the effective action must take the form
\begin{align}
	\begin{split}
		{\cal L}  ={} & \, P(X ) - \frac{1}{4} F_{\mu\nu} F^{\mu\nu}  \\
		& + \frac{1}{2} a(X) \spacy F_{\mu\nu} F^{\mu\nu} + \frac{1}{2} b(X) \spacy \partial^\mu \psi \partial^\nu \psi \spacy F_{\rho \mu} F^\rho {}_\nu \, , \label{full action}
	\end{split}
\end{align}
where $a$ and $b$ are two generic functions and, as before, we are neglecting higher derivative corrections.

In this effective action, the first line describes the superfluid's own dynamics and the free dynamics of the electromagnetic field; the second line couples the two. To understand the physical meaning of the $a$ and $b$ functions, it is useful to set  $\psi$ to its background value and look at the  $F_{\mu\nu}$-dependent part of the action, which becomes
\begin{align} \label{eq: L background}
	\begin{split}
		{\cal L}  \to{} & - \frac{1}{4} (1-2 \bar a) F_{\mu\nu} F^{\mu\nu} + \frac{1}{2} \spacy \bar b \spacy \bar\mu^2 F_{i0} F^i {}_0 \\
		={}& \frac{1}{2} \left( 1-2 \bar a - \bar b  \spacy  \bar\mu^2 \right) \mkern-2mu \bm E^{2} - \frac{1}{2} (1-2 \bar a) \spacy \bm B^{2} \, .
	\end{split}
\end{align}
This allows us to match $a$ and $b$ to the superfluid's dielectric constant $\epsilon$, and magnetic permeability $\mu$~\cite{Caputo:2019xum}: 
\begin{align} \label{eq: epsilon and mu}
	\epsilon(\bar \mu) = 1- 2 \bar a - \bar b  \,  \bar\mu^2 \, , \qquad \mu (\bar \mu)= \frac{1}{1-2 \bar a} \,.
\end{align}
It so happens that most substances have negligible response to external magnetic fields, $\mu \simeq 1$, which means $a\simeq 0$. For simplicity and to avoid clutter---as well as the potential confusion between $\bar \mu$ and $\mu$---we will consider this case only, although extending our analysis to accommodate $\mu \neq 1$ is straightforward.

Finally,  the effective action  \eqref{full action} provides a full description of  the leading interactions between a superfluid and the electromagnetic field at low energies---that is, small frequencies and weak electromagnetic fields. For small perturbations about the ground state, we can expand the action in the phonon field $\pi$. 
The leading interactions  appear at linear order in $\pi$,
\begin{align} \label{couplingsss}
	{\cal L}_{\rm int} = - \Delta \epsilon \,\partial_j \pi  \,  F_{i0} F_{ij}  
+  \frac{1}{2} \frac{d \Delta \epsilon}{d \log \bar \mu} \, \dot \pi \,  F_{i0} F_{i0} \, ,
\end{align}
where we used that the dependence of $b$ on the phonon field is completely fixed by Eq.~\eqref{eq: epsilon and mu}, by simply promoting the background chemical potential to the local one, $\bar\mu \to \sqrt{X}$:
\begin{align}
	b(X) = - X \Delta \epsilon(X) \,, \qquad \text{ with } \; \Delta\epsilon \equiv \epsilon - 1 \, .
\end{align}

Up to now we have kept all relativistic effects. For simplicity, we now take the nonrelativistic limit for our superfluid, in the sense that we assume that the speed of sound is highly subluminal, $c_s \ll 1$, and that the energy density is dominated by the mass density, $\rho_m$, which is also much larger than the pressure. The relativistic enthalpy density and chemical potential then reduce to $w \simeq \rho_m$ and $\bar \mu = m + \mu_{\rm NR}$, with $\mu_{\rm NR} \ll m$. In this case $d \log \bar \mu \simeq d \mu_{\rm NR}/m = d P/\rho_m = c_s^2 \, d \log \rho_m$.  

We further assume that $\Delta \epsilon$ depends non-negligibly on the density of the material,
\begin{align}
	\frac{d \Delta \epsilon}{d \log \rho_m} \sim \Delta \epsilon \, ,
\end{align}
and  that the speed of light in the material is still relativistic, or at least much higher that the speed of sound. In terms of the index of refraction, this means $1/n_0 = 1/\sqrt{\epsilon} \gg c_s$.
So, for on-shell phonons and photons, the second term in Eq.~\eqref{couplingsss} is enhanced with respect to the first one by a factor of $1/c_s$.

Within all these approximations, the leading terms in the Lagrangian are
\begin{align} \label{coupling}
	{\cal L}  =   \frac12\left[ \dot \pi^2 - c_s^2 \big(\bm \nabla \pi\big)^2  + \bm E^{\,2} - \frac{\bm B^{\,2}}{n_0^2} + \frac{g}{n_0^2} \spacy \dot \pi \spacy  \bm E^{\,2} \right] \, .
\end{align}
In the equation above we have canonically normalized the phonon and photon fields, and introduced the coupling
\begin{align}
	g \equiv \frac{1}{ c_s \sqrt{\rho_m}}   \frac{d \Delta \epsilon}{d \log \rho_m} \, .
\end{align}

We stress that these approximations are not necessary. We could carry out the computations that follow without making them. Nonetheless, they are realistic and considerably improve the readability of our analysis, and so we will make them. 
Notice that $n_0$ is, by construction, the zero frequency limit of the index of refraction.

\vspace{1em}

\begin{figure}[t]
\centering
\begin{tikzpicture}[line width=0.9pt]
  \coordinate (i) at (-2.2,0);
  \coordinate (v) at (0,0);
  \coordinate (o1) at (2.2,0.9);
  \coordinate (o2) at (2.2,-0.9);

  \draw[decorate, decoration={snake,amplitude=1.5pt,segment length=6pt}]
    (i) -- (v) node[midway, above,yshift=2pt] {$(\omega, \bm{k}, \hat{\bm{\varepsilon}})$};

  \draw[decorate, decoration={snake,amplitude=1.5pt,segment length=6pt}]
    (v) -- (o1) node[midway, above,yshift=2pt,rotate=19] {$( \omega ', \bm{k}',  \hat{\bm{\varepsilon}}')$};

  \draw[dashed]
    (v) -- (o2) node[midway, below right,xshift=-17pt,yshift=5pt,rotate=-19] {$(E, \bm{p})$};

  \fill (v) circle (2pt);
\end{tikzpicture}
\caption{\label{decay} Single photon decay into a photon and a phonon.}
\label{fig: photon-phonon-decay}
\end{figure}
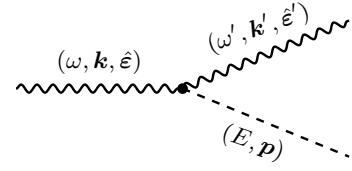

\noindent\emph{\textbf{Photon decay}} ---
Now consider  a photon of momentum $\bm k$ and frequency $\omega = k/n_0$ propagating in the superfluid. The trilinear interaction in the effective action can trigger its decay into a lower energy photon ($\omega'$, $\bm k'$) and a phonon ($E$, $\bm p$), $|\bm k \rangle \ \rightarrow \ |\bm k', \bm p\rangle$ (see Figure~\ref{decay}).
 This is kinematically allowed as long as $1/n_0>c_s$, and can be thought of as Cherenkov emission of phonons.   Working in radiation gauge ($A_0 = \bm \nabla \cdot \bm A = 0$), the amplitude is
\begin{align}
	i {\cal M} =  -\frac{g}{n_0^2} \spacy \omega \spacy E \spacy (\omega-E)\,  \hat{\bm \varepsilon} \cdot \hat{\bm \varepsilon}^{\prime*} \,, 
\end{align}
with $\hat{\bm \varepsilon}$ and $\hat{\bm \varepsilon}^{\,\prime}$ the photons' transverse polarization vectors.

The decay rate can be computed as usual from
\begin{align} \label{eq: single phonon dGamma}
	d \Gamma = \frac{1}{2 \omega} \spacy |{\cal M}|^2 \spacy \frac{d^3 k'}{{(2 \pi)}^3 \, 2 \omega'} \, \frac{d^3 p}{{(2 \pi)}^3 \, 2E} {(2 \pi)}^4 \delta^4({\cdots}) \,,
\end{align}
where the $\delta^4$ enforces energy and momentum conservation. In particular, momentum conservation fixes the momentum of the outgoing photon, yielding $\bm k' = \bm k - \bm p$. After that, energy conservation fixes the angle between $\bm p$ and $\bm k$. In our $n_0 \spacy c_s \ll 1$ approximation, 
\begin{align} \label{phonon angle}
\hat{\bm p}\cdot\hat{\bm k} \ \simeq \ n_0 c_s + \frac{p}{2k}\, .
\end{align}
One recovers the Cherenkov angle in the soft phonon limit $p \ll k$, when the recoil of the source is negligible.  

To zeroth order in $n_0c_s$, $\omega' \simeq \omega$ and $p$ varies between $0$ and $2k$. Moreover, 
the sum over the final photon's polarizations yields a factor
\begin{align}
	\sum_{\hat \varepsilon'} \hat \varepsilon_i \hat \varepsilon_j^* \, \hat \varepsilon_i^{\,\prime *} \hat \varepsilon^{\,\prime}_j 	\simeq 1 - \sin^2(2 \theta) \cos^2 (\phi) \; ,
\end{align}
where $\theta$ and $\phi$ are the polar and azimuthal angles of $\bm p$ in a coordinate system with $\hat{\bm z} = \hat{\bm k}$ and $\hat{\bm x} = \hat{\bm \varepsilon}$. 
Putting everything together and integrating over $\phi$, to lowest order in $n_0 \spacy c_s$ we find the total photon decay rate

\begin{align} \label{total rate}
	\Gamma \simeq \frac{23 \, g^2 c_s}{210 \spacy \pi} \, \omega^5 \, = \, \frac{23 }{210 \spacy \pi} \left(\frac{d \Delta \epsilon}{d \log \rho_m} \right)^2 \frac{\omega^5}{c_s \rho_m}.
\end{align}

This decay process is a form of dissipation, with the photon energy acquiring an imaginary part, $\omega \to \omega - \frac{i}{2} \Gamma$. This must also equal $k/n$, and thus our superfluid's  index of refraction features an imaginary part as well, scaling at low frequencies as $\omega^4$,
\begin{align} \label{eq: Imn}
	{\rm Im} \left(\frac{n(\omega)}{n_0} \right)   \simeq \frac{\Gamma}{2 \spacy \spacy \omega} \simeq \frac{23 \, g^2}{420 \spacy \pi} \spacy c_s \spacy \omega^4 \,,
\end{align}
as well as an associated nonanalyticity in the real part,\footnote{This can be computed from the standard Kramers--Kronig relations on $\epsilon(\omega)$, by isolating the  small-$\omega$ non-analytic behavior of the integral that yields ${\rm Re} \, \epsilon(\omega) $.}
\begin{align}
	{\rm Re} \left( \frac{n(\omega)}{n_0} \right)  &  \simeq 1 + \frac{n_2}{n_0} \, \omega^2 - \frac{23 \spacy g^2 \spacy \omega^4}{420 \spacy \pi^2} \spacy c_s \spacy \log\left( \frac{\omega^2}{\omega^2_0} \right) \, ,
\end{align}
where in both cases we are neglecting powers of $\omega$ higher than four. The constants $n_0$, $n_2$, and $\omega_0$ cannot be computed within the effective theory but can be fixed, in principle, by measuring the diffraction index at three different frequencies. The real prediction of our computation is  the log-running of the coefficient of the $\omega^4$ term.

Aside from overall numerical factors, our result~\eqref{eq: Imn} has exactly the same dependence on frequency and material's properties as Einstein's~\cite{einstein1910theorie}, provided that the temperature is replaced with the typical phonon energy, $T \to E \sim c_s \omega$.  This confirms the picture we alluded to in the introduction, with thermal phonons turning into quantum fluctuations of the phonon field as the temperature approaches zero.

Before moving on, a comment on the regime of validity of our analysis is in order. The effective theory for phonons holds for $p \ll \Lambda$ and $E \ll c_s \Lambda$, with $\Lambda$ of the order of the inverse inter-particle separation. Because the phonon and photon dispersion relations differ, there are now three possible regimes. For $\omega \ll c_s \Lambda \ll \Lambda$, the outgoing phonon has both momentum and energy far below the cutoff. Moreover, multi-phonon processes are subleading, as they carry higher powers of the phonons' energy, which is never larger than $\omega$. Our result~\eqref{total rate} then gives the \emph{total} photon decay rate, up to negligible higher-derivative corrections.
For $c_s \Lambda \lesssim \omega \ll \Lambda$, the outgoing phonon carries $p \sim \omega \ll \Lambda$ and $E \sim c_s \omega \ll c_s \Lambda$, and so the one-phonon rate is still reliably computed within the effective theory. In this regime, however, additional channels (multiple outgoing phonons, new collective modes, and so on) become relevant and add positively to the total rate; our result becomes a sharp (up to higher derivative corrections) \emph{lower bound}.
For $\omega \gtrsim \Lambda$, even the total one-phonon rate becomes incalculable within the effective theory.

\vspace{1em}


\noindent\textbf{\emph{Transparency, Opalescence, Opacity}} --- We now go back to the our main claim, that of an intrinsic limit to transparency. There appear to be two possible notions of the latter, which one can identify respectively as the opposite of \emph{opalescence} or \emph{opacity}.

The decay process described above involves a photon as well as a phonon in the final state. The outgoing phonon takes away a very little fraction of the  photon energy, of order $n_0 c_s  \ll 1$. On the other hand, one can show that photons are typically scattered at large angles in the process. Indeed, most scattered photons are directed \emph{backwards} with respect to the incoming flux.\footnote{Let us go back to the differential rate \eqref{eq: single phonon dGamma}. To lowest order in $n_0 c_s$, the phonon's momentum $p$ ranges between zero and $2 k$, while the outgoing photon's momentum is $k' \simeq k $. Since $\bm{k} = \bm k' + \bm p$, whenever $p >\sqrt{2} \, k$ the outgoing photon is emitted {\em backwards} relative to the beam---that is, with $\hat{\bm k}'  \cdot \hat{\bm k} < 0$. Using the differential rate \eqref{eq: single phonon dGamma}, which is  peaked at high phonon momenta, we find that this is the case for over two thirds of the decay events. This means that the decay process, however rare it might be, when it does happen it tends to drastically change the direction of the photon. }
So photons get diffused in all possible directions much earlier than when they lose an appreciable fraction of the their energy into phonons. The rate at which an incoming beam of photons is {\em scattered} is given by the decay rate already computed in~\eqref{total rate}, because for each phonon produced, as we just discussed, one photon is essentially lost from the beam. This is then the rate at which the material develops opalescence. More concretely, at length scales bigger than $1/(n_0 \Gamma)$, the material is opalescent.

When we say that a medium (say, water) is transparent at certain frequencies over a given distance, we typically mean that objects remain visible through the medium at those distances. In turn, this means that there are enough photons directly emitted by those objects which are not yet scattered in the diffuse background (``into the blue"). This definition of transparency is the opposite of opalescence and is thus characterized by the same length scale $\lesssim 1/(n_0 \Gamma)$. This is also the effect attached to the imaginary part of $n$, as discussed above.

We can instead compute the rate at which the initial beam energy is converted into phonons. This is obtained by integrating the differential rate~\eqref{eq: single phonon dGamma} times the phonon energy and normalizing to the initial energy $\omega$,
\begin{align} \label{eq: h single photon}
\gamma = \frac{1}{\omega} \int E \, d\Gamma = \frac{35}{23} \, n_0 c_s \, \Gamma\, .
\end{align}
At the corresponding length scale,  $1/(n_0\gamma)$, the material becomes opaque, in the sense that also the diffuse light is lost (``out of the blue and into the black"). 

We can also think of opalescence as a loss of linear momentum with a negligible loss of energy. And so, apart from order-one overall factors, $\Gamma$ is the rate at which the momentum of an incoming beam of photons dies off, while their total energy decays at a much slower rate, $\gamma \ll \Gamma$. 

\vspace{1em}


\noindent\textbf{\emph{Nonlinear attenuation of large coherent fields}} --- The computations we have performed so far concerned the decay of a single photon. They thus give the correct attenuation rate for incoherent electromagnetic waves, such as those produced by a thermal source. However, for large coherent electromagnetic fields such as those produced  by a laser, new behaviors arise, potentially leading to a dramatic enhancement of our effect. The reason is that our effect is intrinsically nonlinear. To convince oneself that this is the case, it is enough to consider directly the trilinear coupling in Eq.~\eqref{coupling}: the presence of a classical electromagnetic field acts as a source for the $\pi$ field, creating a $\pi$ proportional to $\bm E^2$. The radiated energy per unit time is thus proportional to $\bm E^4$, that is, to the {\em square} of the electromagnetic energy.

To leading order in the number of photons, an electromagnetic coherent state with a large number of photons can be modeled by a classical field.  Taking  $\bm{E}$ to be a classical field, $\bm{E}_{\rm cl}$, the last interaction term of~\eqref{coupling} becomes a classical source for $\pi$, able to produce phonons from the vacuum, $|0\rangle_\pi \rightarrow |\bm p\rangle_\pi$.
The corresponding event rate is given by
\begin{align} \label{eq: Gamma coh}
	d\Gamma_{\rm cl} ={}& \frac{1}{{\cal T}} \left( \frac{g\spacy E}{2 \spacy n_0^2}\right)^{\!\!2} \frac{d^3p}{(2\pi)^3 2E} \\
	& \!\!\times \left| \int \frac{d\omega}{2\pi} \frac{d^3k}{(2\pi)^3} \tilde{\bm{E}}_{\rm cl}(\omega,\bm k) \cdot \tilde{\bm{E}}_{\rm cl}(E-\omega,\bm p - \bm k) \right|^2 \,, \nonumber
\end{align}
where ${\cal T} \to \infty$ is the total interaction time. 
We will express the classical field profile as a wave packet characterized by a momentum wave-function $f(\bm{k})$,
\begin{align}
	\bm{E}_{\rm cl}(x) = \int \frac{d^3k}{(2\pi)^3} \sqrt{\frac{\omega}{2}} \bigg[ \hat{\bm\varepsilon}_{\bm k} \spacy f(\bm k) \spacy e^{-i \omega t + i \bm{k}\cdot\bm{x}} + {\rm h.c.} \bigg] \,.
\end{align}

 An important feature of the phonon emission rate \eqref{eq: Gamma coh} is that it vanishes in the monochromatic plane-wave limit, and is thus controlled by the momentum spread of the incoming electromagnetic wave. This can be understood directly from the convolution in the second line which, after imposing energy conservation, only involves cross terms with positive- and negative-frequency components of the electric field,  $f(\bm k) f^*(\bm k - \bm p)$ and $f(\bm p - \bm k) f^*(-\bm k)$. Emission of a phonon with nonzero momentum \(\bm p\) requires the two factors to have overlapping support, which in turn requires \(f\) to have a finite width.

To make everything concrete, we specialize to a Gaussian wave-packet peaked around some momentum $\bm{k}_\star$,
\begin{align}
	f(\bm k) \propto \exp\!\left(-\frac{(k_\parallel - k_\star)^2}{2\Delta k_\parallel^2} - \frac{\bm k_\perp^2}{2\Delta k_\perp^2}\right) \,,
\end{align}
with the overall normalization related to the average number of photons by
\begin{align} \label{eq: N normalization}
	N_\gamma = \int \frac{d^3k}{(2\pi)^3} \big| f(\bm k) \big|^2 \,.
\end{align}
 This corresponds to a classical field profile with typical wavelength $2\pi/k_\star$, longitudinal spatial extension of order $1/\Delta k_\parallel$, and transverse one of order $1/\Delta k_\perp$. Our intent is to model a laser pulse, and we thus assume $k_\star \gg \Delta k_{\parallel,\perp}$ and $\Delta k_\perp \gg \Delta k_\parallel$. For this configuration, the polarization vectors in the wave-packet all point approximately in the same direction, and so we can set $\hat{\bm\varepsilon}_{\bm k}\cdot \hat{\bm\varepsilon}_{\bm k-\bm p}^{*} \simeq 1$. 
 
The attenuation rate of a classical field is measured by the emitted power $P$ divided by the energy of the incoming electromagnetic wave,
\begin{align}
\gamma_{\rm cl} = \frac{P}{E_{\rm in}} \, ,
\end{align}
where $E_{\rm in} \simeq N_\gamma \, \omega_\star$. This is the generalization of Eq.~\eqref{eq: h single photon} to a classical field. The emitted power can be computed from our phonon emission rate above by simply assigning energy $c_s p$ to each emitted phonon, $dP = c_s p \, d\Gamma_{\rm cl}$.
Upon integrating, we find the attenuation rate
 \begin{align} \label{att2}
\gamma_{\rm cl} = \frac{N_\gamma g^2 c_s^2}{ 8 \pi n_0^3} \frac{\omega_\star}{L_{\rm eff}^4} = \frac{3N_\gamma}{4} \left( \frac{1}{L_{\rm eff} k_\star} \right)^{\!4} \gamma_\star \, , 
\end{align}
where $\gamma_\star$ is the corresponding quantum version, Eq.~\eqref{eq: h single photon}, evaluated at the beam momentum $k_\star$. We also defined the effective beam size:
\begin{align}
L_{\rm eff}^2 \equiv \frac1{\Delta k_\perp^2} + \left(\frac{n_0 c_s}{\Delta k_\parallel}\right)^{\!2} \, .
\end{align}
  
Notice that, in the limit in which we can treat the electromagnetic coherent state as a classical field, the opalescence effect described above for incoherent photons is absent. To show this, we notice that for our wave packet, which is sharply peaked in momentum space, the outgoing phonons are emitted along the Mach cone, with $\hat {\bm p} \cdot \hat {\bm k}_\star \simeq n_0 c_s $. Then, an emitted phonon of energy $E$ carries away a momentum $\bm p$ whose magnitude is $E/c_s$, but whose {\em longitudinal} component is  $n_0 E$. Averaging over the azimuthal angle around the beam, we see that the emitted phonons take away a total momentum that is $n_0$ times the energy they take away, and aligned with the beam. But the beam itself has a total momentum that is $n_0$ times its energy, which means that its momentum and energy decay at the same rate, with no room for opalescence.

\vspace{1em}


\noindent\textbf{\emph{Estimates for helium-4}} --- How stringent is this fundamental limit to transparency? Let's consider helium-4 as a benchmark superfluid. At zero temperature and saturated vapor pressure, its speed of sound and mass density are $c_s \simeq 238 \text{ m/s}$ and $\rho_m \simeq 0.14 \text{ g/cm}^3$~\cite[e.g.,][]{donnelly1998observed}. Using the Clausius--Mossotti relation~\cite[e.g.,][]{donnelly1998observed} one also finds $d\Delta\epsilon/d\log\rho_m \simeq \Delta \epsilon \simeq 0.057$. The momentum cutoff $\Lambda$ can be estimated as that momentum for which the phonon dispersion relation starts deviating appreciably from linearity---the hallmark that higher derivative corrections become relevant. By doing that, one gets $\Lambda \simeq 1.5 \text{ keV}$~\cite[e.g.,][]{Esposito:2026diu}. Given this, we can estimate the typical opalescence length of incoherent photons, $1/(n_0 \Gamma)$, with $\Gamma$ the single photon rate~\eqref{total rate}. As discussed, depending on whether the photon energy if below or above $\Lambda$, this provides either the length itself, or an upper bound to it. With this in mind, we get
\begin{subequations}
\begin{align}
	1/(n_0 \Gamma) ={}& 8 \times 10^{28} \text{ m} \qquad \text{ for } \omega = 0.1 \text{ meV} \,, \\
	1/(n_0 \Gamma) \leq{}& 8 \text{ cm} \qquad\qquad\;\,\, \text{ for } \omega = 0.1 \text{ keV} \,.
\end{align}
\end{subequations}
Owing to the derivative nature of the Goldstone's interactions, a low-energy photon decays very rarely, and the corresponding opalescence length is very large. In fact, at $\omega=0.1 \text{ meV}$, larger than the observable size of the Universe. Nonetheless, the decay rate~\eqref{total rate} depends strongly on frequency, and the opalescence length for a soft X-ray photon of $\omega = 0.1 \text{ keV}$ is less than 8~cm. Yet, photons of this energy can excite several different degrees of freedom, besides quantum phonons. Most notably, their energy is above the photoionization threshold, which we expect to dominate the decay rate. Indeed, the measured attenuation length at this energy in helium-4 is $862 \text{ }\mu\text{m}$~\cite{henke1993x}. Our estimates so far have been for the opalescence length. Given that $\gamma \sim c_s \Gamma \ll \Gamma$, the corresponding opacity length is about a million times larger.

How about coherent effects? For how we formulated the problem so far, the natural comparison is between the opacity length of an incoherent beam $1/\gamma$, and that of a coherent classical one, $1/\gamma_{\rm cl}$. Whether or not coherence can reduce the opacity length with respect to the incoherent case depends on the competition between two effects in \eqref{att2}. On the one hand, $\gamma_{\rm cl}$ increases with increasing number of average photons, $N_\gamma \gg 1$. On the other, it decreases with increasing beam size, $L_{\rm eff} k_\star \gg 1$. Which of these two effects dominates is a matter of which laser one considers.

A laser pulse of energy $E_{\rm in}$ contains an average number of photons $N_\gamma \simeq E_{\rm in}/\omega_\star$. The  longitudinal and orthogonal pulse sizes are of the order $L_{\parallel, \perp} \equiv 1/\Delta k_{\parallel,\perp}$. 
Let's start considering a Q-switched ruby laser, with wavelength $2\pi/k_\star \simeq 694 \text{ nm}$. The typical peak power of such a laser if $P_{\rm peak} \sim 10 \text{ MW}$, and the typical pulse duration is of order $\tau \sim 10 \text{ ns}$~\cite[e.g.,][]{svelto1998principles}, which implies an average number of photons $N_\gamma \simeq E_{\rm in}/\omega_\star \simeq P_{\rm peak} \, \tau / \omega_\star \sim 3 \times 10^{17}$. The typical transverse size of the laser is set by the diameter of the ruby rod itself, giving $L_\perp \sim 1 \text{ cm}$~\cite[e.g.,][]{svelto1998principles}. Since $L_\parallel \sim \tau$, for the value of $c_s$ reported above, we also find $L_{\rm eff} \simeq L_\perp$. The second benchmark we consider is a pulsed CO$_2$ laser, with a wavelength of $10 \text{ }\mu\text{m}$, a pulse duration of order $50 \text{ ns}$, and a pulse energy $E_{\rm in} \simeq 3.8 \text{ J}$~\cite{zhi2010efficient}, leading to $N_\gamma \simeq 2 \times 10^{20}$. The transverse size is again of order $1 \text{ cm}$~\cite[e.g.,][]{svelto1998principles}, and thus again $L_{\rm eff} \simeq L_\perp$. With this at hand, we find the following relation between the classical and quantum opacity lengths,
\begin{subequations}	
\begin{align}
	\frac{1}{\gamma_{\rm cl}} \sim{}& 300 \, \frac{1}{\gamma} \qquad\;\; \text{ for ruby} \,, \\
	\frac{1}{\gamma_{\rm cl}} \sim{}& 10^{-5} \, \frac{1}{\gamma} \qquad \text{ for CO}_2 \,.
\end{align}
\end{subequations}
Hence, we can find instances where coherence effects dominate the opacity length. Nonetheless, at least for the example above, the quantum opacity length is so large to lead to a still very large classical one, of order $10^{19} \text{ m}$.

Whether or not there are systems for which the fundamental quantum bounds we derive are phenomenologically relevant is, for the moment, an open question.

\vspace{1em}


\begin{acknowledgments}
\noindent\textbf{\emph{Acknowledgments}} --- We thank Ana Asenjo-Garcia, Yuri Levin, and Stefano Liberati for their insightful questions and comments.  FP received support from the French government under the France 2030 investment plan, as part of the Initiative d'Excellence d'Aix-Marseille Universit\'e - A*MIDEX (AMX-19-IET-012) and from the ``action th\'ematique" Cosmology-Galaxies (ATCG) of the CNRS/INSU PN Astro and by the {\it Agence Nationale de la Recherche} under the grant ANR-24-CE31-6963-01. AN is partially supported by the U.S.~Department of Energy (award number DE-SC0011941).
\end{acknowledgments}

\bibliographystyle{apsrev4-1}
\bibliography{biblio}

\end{document}